\documentclass[twocolumn,english,aps,prb,twocolum,superscriptaddress,bibnotes,amsmath,amssymb,floatfix]{revtex4-2}
\usepackage[colorlinks=true,citecolor=blue,linkcolor=magenta]{hyperref}

\usepackage[markup=nocolor, authormarkupposition=left]{changes} 
\usepackage{soul}
\usepackage[utf8]{inputenc}
\usepackage[english]{babel}
\usepackage{amsmath,amsfonts,amssymb}
\usepackage[T1]{fontenc}
\usepackage{url}
\usepackage{amsmath}
\usepackage{amsfonts}
\usepackage{amssymb}
\usepackage{epstopdf}
\usepackage{graphicx}
\DeclareUnicodeCharacter{2009}{ }

\begin{document}
\title{Broadband quantum-dot frequency-modulated comb laser}

\author{
Bozhang Dong$^{1,\ast, \dagger}$,
Mario Dumont$^{2, \ast}$,
Osama Terra$^{1}$,
Heming Wang$^{1}$,
Andrew Netherton$^{2}$,
and John E. Bowers$^{1,2,\dagger}$\\
\textit{
$^1$Institute for Energy Efficiency, University of California, Santa Barbara, CA, USA\\
$^2$Department of Electrical and Computer Engineering, University of California, Santa Barbara, CA, USA\\}
$^\dagger$Email: bdong@ucsb.edu, bowers@ece.ucsb.edu\\
$^*$These authors contributed equally\\
}

\begin{abstract}
Frequency-modulated (FM) laser combs, which offer a periodic quasi-continuous-wave output and a flat-topped optical spectrum, are emerging as a promising solution for wavelength-division multiplexing applications, precision metrology, and ultrafast optical ranging. The generation of FM combs relies on spatial hole burning, group velocity dispersion (GVD), Kerr nonlinearity, and four-wave mixing (FWM). While FM combs have been widely observed in quantum cascade Fabry-Perot (FP) lasers, the requirement for a low-dispersion FP cavity can be a challenge in platforms where the waveguide dispersion is mainly determined by the material. Here we report a 60 GHz quantum-dot (QD) mode-locked laser in which both the amplitude-modulated (AM) and the FM comb can be generated independently. The high FWM efficiency of -5 dB allows the QD laser to generate an FM comb efficiently. We also demonstrate that the Kerr nonlinearity can be practically engineered to improve the FM comb bandwidth without the need for GVD engineering. The maximum 3-dB bandwidth that our QD platform can deliver is as large as 2.2 THz. This study gives novel insights into the improvement of FM combs and paves the way for small-footprint, electrically-pumped, and energy-efficient frequency combs for silicon photonic integrated circuits (PICs).

\end{abstract}
\maketitle

\medskip
\begin{large}
\noindent \textbf{Introduction} 
\end{large}

Since the conception of the laser frequency comb in the late 1990s, it has revolutionized the precise measurement of frequency and time. An optical frequency comb (OFC) refers to a coherent source composed of discrete and equally spaced frequency lines in the optical domain. In the time domain, the emission of an OFC is a periodic signal before their eventual phase drift due to the interaction with the environment. Beyond their initial use in optical clocks and precision spectroscopy \cite{Picque:19,terra2019absolute}, frequency combs have exhibited strong potential for various emerging applications, including ultraviolet and infrared (IR) spectroscopy, remote sensing, optical frequency synthesis, and broadband dense wavelength-division multiplexing (DWDM) systems \cite{yu2018silicon,spencer2018optical,corcoran2020ultra,diddams2020optical}. Recently, the development of integrated OFC technologies allows for lowering the system size, weight, power consumption, and cost (SWaP-C) \cite{chang2022integrated}, which enables the OFC technologies to be commercialized in autonomous driving, 5G/6G communications, and artificial intelligence (AI) \cite{riemensberger2020massively,feldmann2021parallel}. During the past decade, frequency-modulated (FM) laser comb has experienced a renaissance in both quantum cascade lasers (QCLs) and quantum-dot (QD) lasers and is currently regarded as a promising solution for technologies requiring OFCs. In contrast to conventional amplitude-modulated (AM) mode-locked lasers where the comb lines are in phase, FM mode-locked lasers obey a periodic modulation of the output frequency, which corresponds to a splay-phase state. In the time domain, the emission of an FM comb laser is a series of quasi-continuous-wave (quasi-CW) periodic short pulses. Soon after the experimental observation of FM combs, they became a promising solution for broadband optical communication. For instance, their quasi-CW operation is beneficial for minimizing the detrimental phase effect caused by high-intensity intra-cavity pulse propagation. For photonic integrated circuits (PICs) where microring modulator arrays are deployed, the high instantaneous power of a conventional AM comb source would result in strong thermal nonlinearities potentially leading to a degradation of data transmission \cite{de2019power}. Also, FM combs are known for broadband and flat-topped ("top-hat-like") profiles in the optical domain, which is crucial in DWDM applications since all comb lines are usable for data transmission because of their comparable powers.

\begin{figure*}[t!]
\centering
\includegraphics[width=\linewidth]{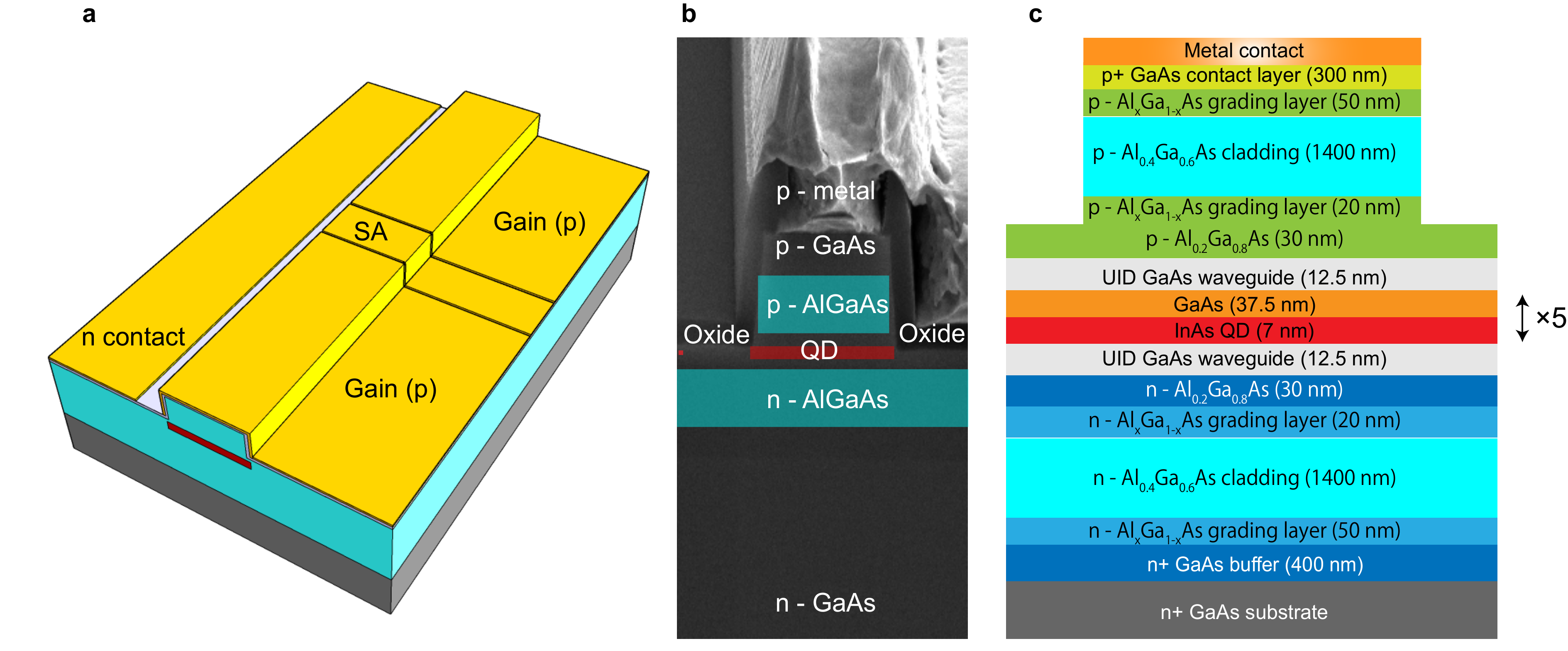}
\caption{\textbf{Design of the 60 GHz QD mode-locked laser.}
\textbf{a}. Schematic of the InAs/GaAs colliding-pulse QD laser. The total cavity length is 1.5 mm, and the saturable absorber whose length is 7.8\% of the total cavity length is placed at the center of the cavity.
\textbf{b}. Cross-section SEM image of the shallow-etched FP cavity.
\textbf{c}. Schematic diagram of the epitaxial structure. 
}
\label{Fig:SEM}
\end{figure*}

The generation of FM combs does not only rely on the group velocity dispersion (GVD) of the waveguide, but also the nonlinear properties of laser active region, including spatial hole burning (SpaHB), Kerr nonlinearities, and four-wave mixing (FWM). In both theory and experiment, FM combs have been demonstrated in mature semiconductor laser systems, including quantum-well (QW) \cite{calo2015single,dong2018physics}, QD/quantum-dash (QDash) \cite{rosales2012high,gioannini2015time}, QCLs \cite{hugi2012mid,hillbrand2019coherent}, and interband cascade lasers (ICLs) \cite{schwarz2019monolithic}. It should be noted that the QD laser offers a unique opportunity for the generation of both AM and FM combs \cite{hillbrand2020phase}. Depending on the injection current, the QD active region is switched from an interband slow gain to an intersubband fast gain. The latter is attributed to the carrier filling up in the higher excited states that allows for an ultrashort intersubband carrier recovery time of 100 fs \cite{borri2000spectral}. On the other hand, the QD laser is also an ideal on-chip light source for silicon PICs, owing to their three-dimensional charge carrier confinement which facilitates their immunity to the growth defects that arise during epitaxial growth of III-Vs on Si. This, along with ultra-low threshold currents, high-temperature operation, very long device lifetimes, high scalability with 300 mm Si wafers, and a high tolerance for chip-scale back-reflection \cite{shang2021perspectives,grillot2021uncovering,dong2021dynamic,shang2022electrically} ensures QD lasers will play a crucial role in the next generation of PICs and integrated OFC technologies. During the past decade, both the single-section QD FM comb and the passively mode-locked QD AM comb have been demonstrated \cite{liu2019high,pan2020quantum,huang2022ultra,liang2022energy}. Nevertheless, broadband and energy-efficient coexisting AM/FM QD mode-locked laser was barely reported. The understanding of AM and FM natures, in particular, their connection and distinction in QD lasers, remains to be explored and improved for their use in industry.

In this work, we demonstrate a high-performance O-band QD mode-locked laser in which the AM and the FM comb can be generated independently. A colliding-pulse structure allows the laser to have a fast repetition rate of 60 GHz, which is beneficial to DWDM applications. Kerr nonlinearity, FWM, and GVD of the waveguide are investigated to explain how to distinguish and generate the AM and the FM comb in this type of device. In particular, we reveal that the Kerr nonlinearity and the FWM in the QD laser can be practically engineered to improve the FM comb bandwidth, which gives novel insights into the development of the FM comb laser.

\begin{figure}[t!]
\centering
\includegraphics[width=\linewidth]{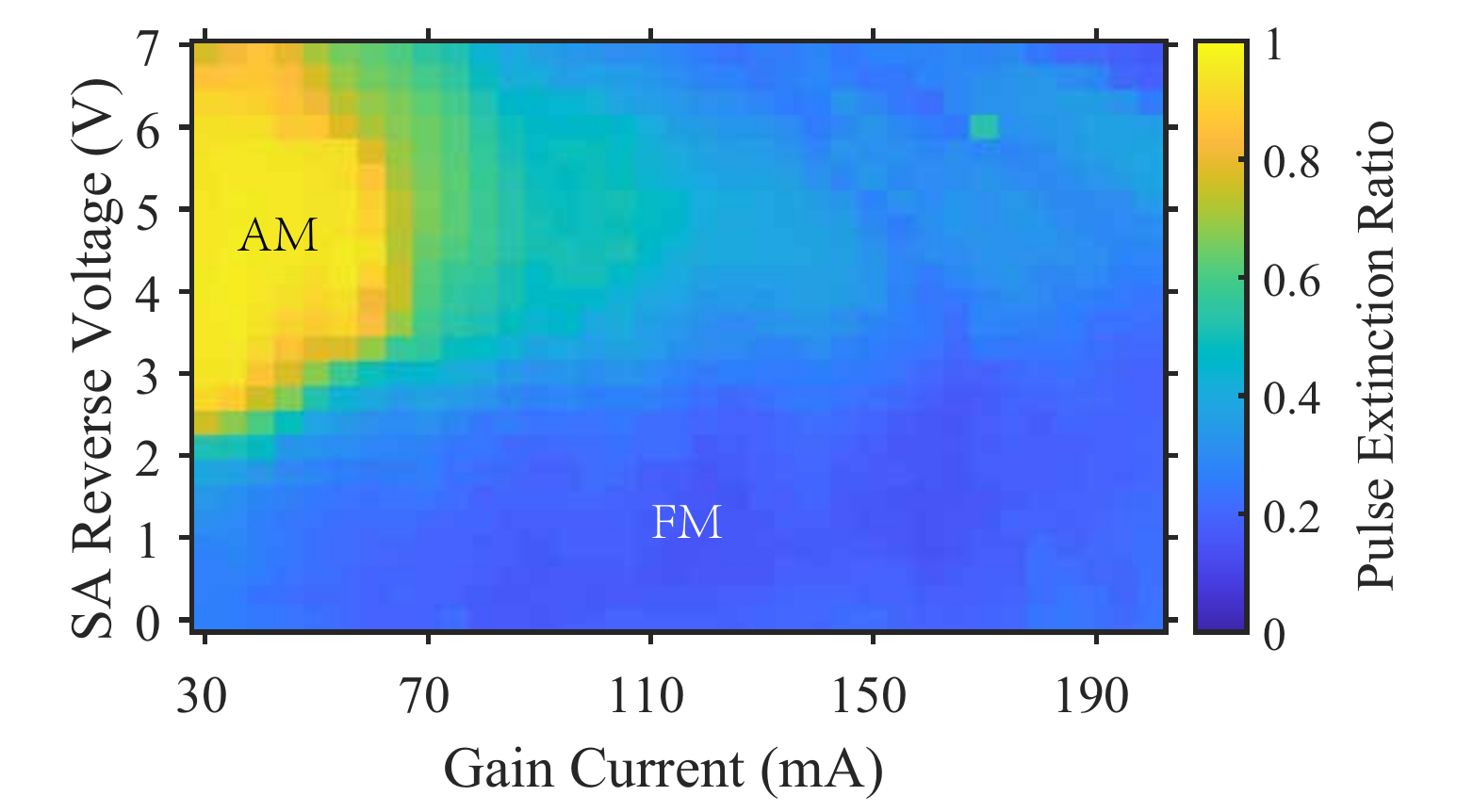}
\caption{\textbf{Pulse extinction ratio in the QD mode-locked laser.}
The formation of an AM comb requires that the gain recovery is slower than the SA recovery. The QD laser should operate with a low gain current and a high SA reverse bias to generate an AM comb. An increase in the gain current leads to the formation of an FM comb owing to the four-wave mixing.
}
\label{Fig:AM_FM}
\end{figure}

\begin{figure*}[t!]
\centering
\includegraphics[width=\linewidth]{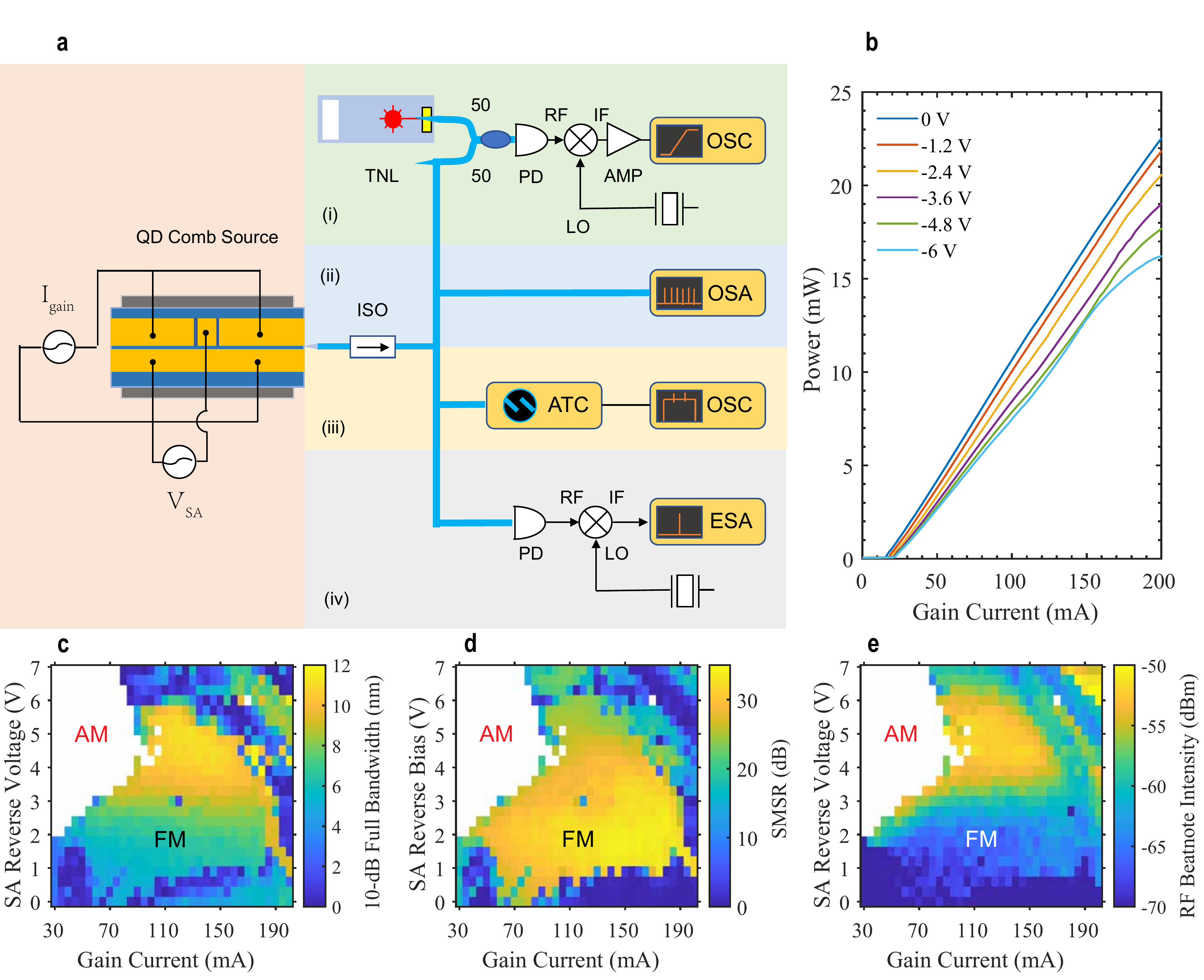}
\caption{\textbf{Characterizations of 60 GHz QD FM comb laser.}
\textbf{a}. Experimental setup for FM comb characterization, including (i) stepped heterodyne method for spectral phase measurements; (ii) optical spectra measurements; (iii) temporal pulse train measurements with autocorrelation; (iv) RF beatnote measurements. ISO, optical isolator; TNL, tunable laser; PD, photodiode; AMP, RF amplifier; OSC, oscilloscope; OSA, optical spectrum analyzer; ATC, autocorrelator; ESA, electrical spectrum analyzer. RF, radio frequency; LO, local oscillator; IF, intermediate frequency.
\textbf{b}. Light-current (L-I) curves of the QD laser under different reverse biases on the saturable absorber (SA) at 20$^{\circ}$C.
\textbf{c}. Mapping of 10-dB optical bandwidth under different biases on the gain and the SA.
\textbf{d}. Mapping of average side-mode-suppression-ratio (SMSR) among the central 20 comb lines under different biases on the gain and the SA.
\textbf{e}. Mapping of RF beatnote intensity under different biases on the gain and the SA. AM, amplitude-modulated dominant operation; FM, frequency-modulated dominant operation.
}
\label{Fig:1}
\end{figure*}
\begin{figure*}[t!]
\centering
\includegraphics[width=0.95\linewidth]{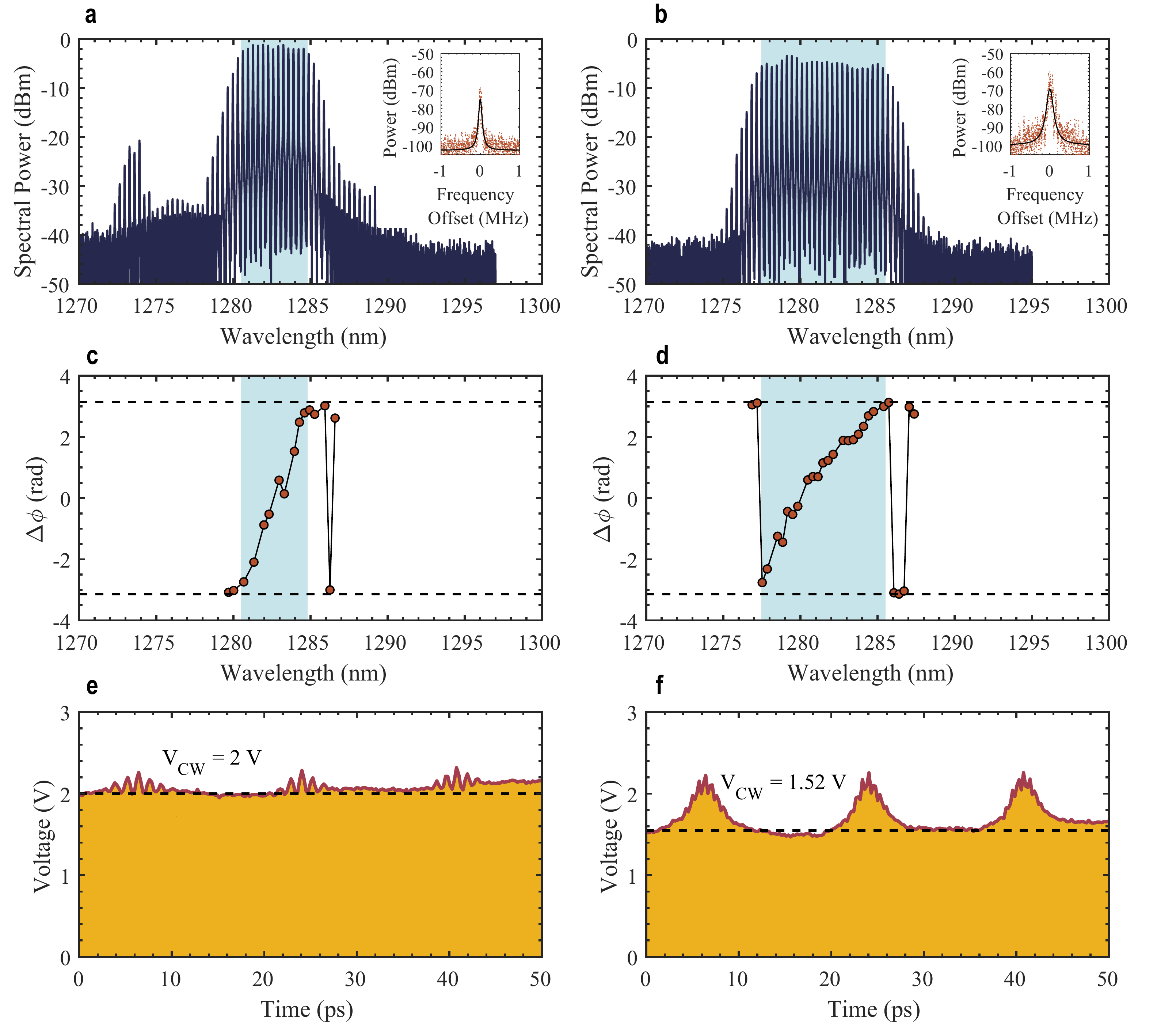}
\caption{\textbf{FM comb dynamics in the QD laser.}
\textbf{a}. Optical spectrum and inset the RF beatnote after it is down-converted to 20 GHz.
\textbf{c}. Spectral intermode phase difference.
\textbf{e}. Intensity autocorrelation in time. The operation conditions for the left column is ($I_{gain}$, $V_{SA}$)=(110 mA, -0.9 V).
\textbf{b}. Optical spectrum and inset the RF beatnote after it is down-converted to 20 GHz.
\textbf{d}. Spectral intermode phase difference.
\textbf{f}. Intensity autocorrelation in time. The operation conditions for the right column is ($I_{gain}$, $V_{SA}$)=(110 mA, -3.3 V).
The FM comb bands are highlighted in cyan.
}
\label{Fig:2}
\end{figure*}

\medskip
\begin{large}
\noindent \textbf{Results} 
\end{large}

Figure \ref{Fig:SEM}a schematically presents the structure of the QD mode-locked laser. The device was grown with a Varian GenII Molecular Beam Epitaxy chamber on nominal 001 n++ GaAs substrates. The SEN image and the epi structure are shown in Fig. \ref{Fig:SEM}b and c, respectively (Methods). The total cavity length is designed to be 1.35 mm, which produces a fundamental cavity resonance at 30 GHz. 
By placing an SA made from the same material as the gain at the center of the laser cavity, the QD laser can take advantage of the second harmonic of the cavity resonance, and the mode spacing is doubled to 60 GHz for DWDM applications. A short SA whose length is 7.8\% of the total cavity length strikes a good balance between efficiency and performance. The FP cavity is shallow-etched and the ridge width is reduced to 2.6 $\mu m$ to improve the wall-plug efficiency (WPE) over 10\% (single-side) (Fig. \ref{Fig:SEM}(a)). Both facets of the FP cavity are left as-cleaved to enable an efficient laser emission. The formation of the AM/FM comb is dependent on the bias conditions on the gain and the SA sections. Figure \ref{Fig:1}b depicts the light-current (L-I) curves under different reverse biases on the SA. With the increase in SA reverse voltage from 0 to -6 V, the slight increase in the internal loss results in an increase in the threshold current from 15 to 21 mA along with a decrease in single-facet external efficiency $\eta$ from 13\% to 10.1\%. The latter is calculated by $\eta=\frac{q\lambda}{hc} \frac{\Delta P}{\Delta I}$, with $h$ the Planck constant, $c$ the speed of light, $q$ the electron charge, and $\lambda$ the lasing wavelength. In all SA bias conditions, the QD laser studied exhibits a sufficient free-space output power over 15 mW. 

\begin{figure*}[t!]
\centering
\includegraphics[width=\linewidth]{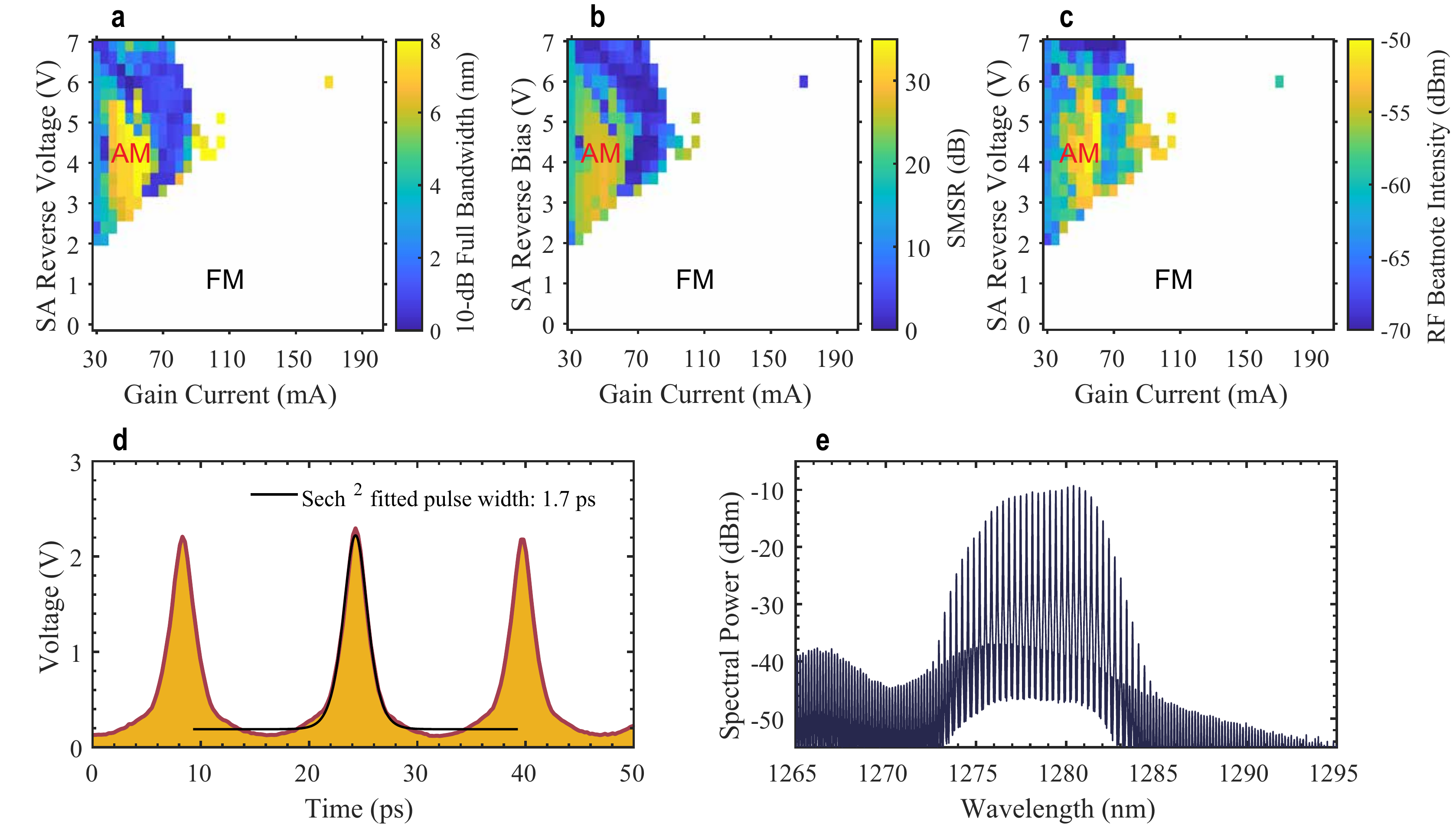}
\caption{
\textbf{AM comb dynamics in the QD laser}. 
\textbf{a}. Mapping of the 10-dB optical bandwidth under different biases on the gain and the SA.
\textbf{b}. Mapping of the average side-mode-suppression-ratio (SMSR) among the central 20 comb lines under different biases on the gain and the SA.
\textbf{c}. Mapping of the RF beatnote intensity under different biases on the gain and the SA. AM, amplitude-modulated dominant operation; FM, frequency-modulated dominant operation.
\textbf{d}. Intensity autocorrelation of an AM comb state. 
\textbf{e}. Optical spectrum of an AM comb state. The gain current is 45 mA and the SA reverse voltage is -4.5 V.}
\label{Fig:SI_AM}
\end{figure*}

\medskip
\noindent \textbf{Generation of FM comb}

The generation of an FM comb depends on a fast gain. Applying a high injection current to the QD gain section allows for fast intersubband carrier capture and relaxation. In this condition, the SA recovery rate is too slow to open a net gain window for the AM pulse generation, and the FM operation is dominant. To determine whether a state is AM or FM operation, here we define a simplified pulse extinction ratio $\eta_p$. The experimental setup for optical pulse characterizations is shown in Fig. \ref{Fig:1}a(iii) (Methods). An example of the intensity autocorrelation for the FM comb emission in time is shown in Fig. \ref{Fig:2}f. By using the CW power level $V_{CW}$ and the peak pulse power level $V_{p}$, the pulse extinction ratio is then expressed as follows: 
\begin{equation}
  \eta_{p} = 1-\frac{V_{CW}}{V_p} 
  \begin{cases}
      \ge 0.5 & \text{(AM domination)}\\
      < 0.5 & \text{(FM domination)}
    \end{cases}   
\label{equ:AMFM}
\end{equation}
Figure \ref{Fig:AM_FM} depicts the pulse extinction ratio of the QD mode-locked laser in different operation conditions. It should be noted that the definition of $\eta_p$ = 0.5 is a rough estimation for the boundary of AM and FM operation. Pure AM state take place when $\eta_p$ close to 1. It should be noted that $\eta_p =$ 0 is not a necessary condition for an FM comb since an FM comb also produces short pulses due to the Kerr nonlinearity. In this work, we demonstrate that $\eta_p =$ 0.34 ensures an FM comb. Therefore, either AM or FM comb plays a dominant role in the laser's operation while $0<\eta_p <1$. Following the definition of Eq. (\ref{equ:AMFM}), the 10-dB optical bandwidth, the average SMSR, and the RF beatnote intensity in FM operation are shown in Figs. \ref{Fig:1}c-e, respectively. The experimental setups corresponding to each measurement are shown in Fig. \ref{Fig:1}a (Methods). 

The FM comb dynamics are summarized in Figs. \ref{Fig:2}. The optical spectra are shown in the first row, where the insets depict the RF spectra for the beatnote after it is down-converted to 20 GHz. The second row depicts the spectral phase difference between adjacent comb lines. The timing traces obtained by intensity autocorrelation are then shown in the third row. In this study, the intermode phase differences $\Delta\phi$ are measured by using the stepped heterodyne method; the corresponding experimental setup is shown in Fig. \ref{Fig:1}a(i) (Methods). A low SA reverse bias allows for suppressing the AM operation efficiently. The left column in Figs. \ref{Fig:2} summarizes the laser dynamics when it operates with a gain current of 110 mA and a low SA reverse bias of -0.9 V. The QD laser operates in the FM mode in the presence of the intermode beating phases uniformly distributed over the range of 2$\pi$ (Fig. \ref{Fig:2}c). As a result, the laser delivers a series of short FM pulses with $\eta_p$ = 0.13 (Fig. \ref{Fig:2}e). The FM mode also enables the laser a flat-topped coherent comb operation in the presence of the 10-dB optical bandwidth of 5.6 nm (1.02 THz) and a narrow RF beatnote linewidth less than 20 kHz (Fig. \ref{Fig:2}a and the inset). An increase in the SA reverse voltage to -3.3 V gives rise to an increase in the SA recovery rate and the Kerr nonlinearity. The consequent increase in the AM pulse strength results in an increase in the $\eta_p$ to 0.34 (Fig. \ref{Fig:2}f). It also contributes to an increase in the RF beatnote intensity by 10 dB (Inset in Fig. \ref{Fig:2}b). However, the laser still operates in the FM mode in the presence of the intermode beating phases uniformly distributed over the range of 2$\pi$ (Fig. \ref{Fig:2}d). It should be noted that the SA reverse bias contributes to an improvement in the FM comb bandwidth. For instance, the 3-dB optical bandwidth of the QD laser, which is mainly determined by the FM bandwidth, is improved from 4.2 nm (0.78 THz) to 8.2 nm (1.49 THz) as the SA reverse bias increases from -0.9 to -3.3 V. Meanwhile, a high SMSR over 30 dB is maintained in the optimum FM operation regime. Nevertheless, the RF linewidth is broadened to 40 kHz from the additional frequency drift of the newly generated comb lines. In this study, the shortest FM pulse width of 495 fs is obtained when the QD laser operates with 55 mA applied to the gain and 0V applied to the SA (Supplementary Information). The optimal FM comb bandwidth state is found when the QD laser operates with 135 mA applied to the gain and -4.8 V applied to the SA, where the 3-dB and the 10-dB comb bandwidth are further improved to 10 nm (1.82 THz) and 11.5 nm (2.12 THz), respectively. The maximum 3-dB comb bandwidth that our device can deliver is 12.1 nm (2.2 THz) after optimizing the cavity design. It should be noted that there is no evident degradation in the comb bandwidth when the operation temperature is increased to 45$^{\circ}$C (Supplementary Information).

\medskip
\noindent \textbf{Generation of AM comb}

The QD is switched back to a slow gain medium when the injection current is low, which allows the SA recovery rate to surpass the gain recovery rate. As a result, a periodic net gain window is opened and the laser operates in the AM mode. An increase in the SA reverse bias is beneficial to the AM comb generation. The AM comb dynamics are summarized in Figs. \ref{Fig:SI_AM}. The QD laser suffers from a reduction in the comb bandwidth in AM operation due to the decrease in the SpaHB. On the other hand, the average SMSR of the AM comb is also lower than the FM comb, due to the Gaussian-shape distribution of modal gains. An example of the AM comb is shown in Fig. \ref{Fig:SI_AM}e, where the QD laser operates with 45 mA applied to the gain and -4.5 V applied to the SA. Corresponding intensity autocorrelation is shown in Fig. \ref{Fig:SI_AM}d. In the AM-dominant comb regime, the pulse width ranges from 1.7 to 3 ps by assuming a sech$^2$ shape, and the RF beatnote intensity is as large as -50 dBm.

\medskip
\noindent \textbf{FM comb formation with Kerr nonlinearity}

\begin{figure}[t!]
\centering
\includegraphics[width=\linewidth]{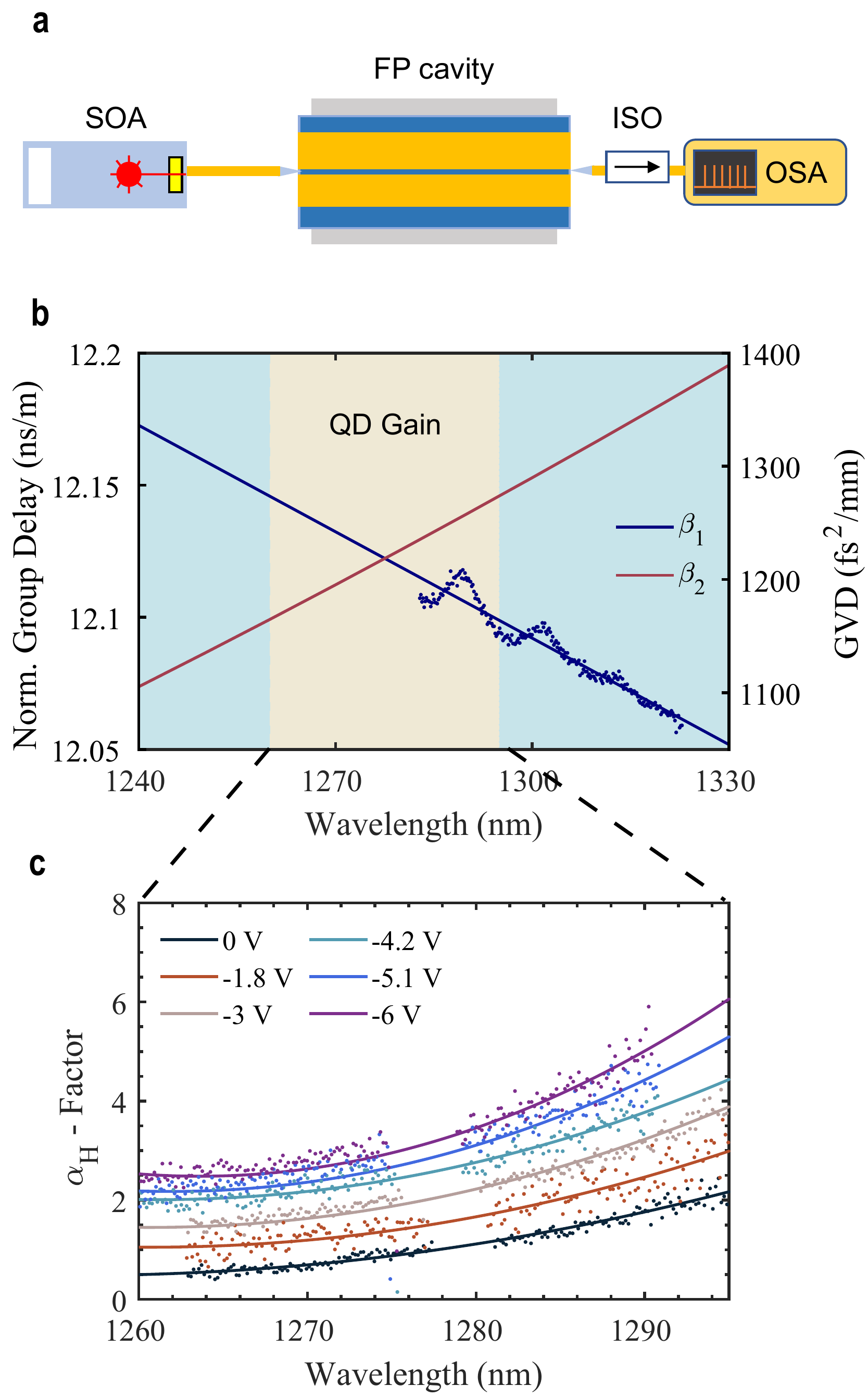}
\caption{\textbf{Influence of SA reverse bias on the Kerr nonlinearity.}
\textbf{a}. Experimental setup for the measurements of group velocity dispersion (GVD). SOA, wideband semiconductor optical amplifier; ISO, optical isolator; OSA, optical spectrum analyzer. 
\textbf{b}. Measured normalized group delay (blue) and calculated GVD (red) as a function of wavelength. 
\textbf{c}. Wavelength-dependent linewidth enhancement factors under different SA reverse biases.
}
\label{Fig:3}
\end{figure}

The improvement of FM comb bandwidth as the SA reverse bias increases from -0.9 to -3.3 V is attributed to the Kerr nonlinearity. In single-section QD lasers or QCLs, the Kerr effect, which acts as an equivalent, lumped, and ultrafast SA, plays an important role in self-mode-locking by triggering multimode instabilities \cite{gordon2008multimode,jiao2013modeling}. As such, an extra SA section in the laser cavity can further contribute to a greater Kerr nonlinearity. The enhanced Kerr effect in III-V lasers originates from a change in the real part of the refractive index caused by a change in the population distribution in the laser cavity. As the intracavity field intensity increases, the mode is more confined in the active region, which results in an increase in the modal gain along with a decrease in the loss. Since the modal gain is tightly coupled to the light intensity, the Kerr effect becomes significant in a fast gain medium with a non-zero $\alpha_H$-factor, which is expressed as follows \cite{osinski1987linewidth}:
\begin{equation}
    \alpha_H = -2\frac{\omega}{c} \frac{dn/dN}{dg/dN} = -\frac{4\pi}{\lambda} \frac{dn/dN}{dg/dN}
    \label{equ:LEF_theory}
\end{equation}
with $N$ the carrier density injected into the laser, $\lambda$ the lasing wavelength, $dn/dN$ the differential refractive index, and $dg/dN$ the differential gain. The $\alpha_H$-factor is thus used to describe the coupling degree between the carrier density-induced gain and refractive index variations. Nevertheless, it should be noted that the Kerr effect is much weaker in lasers with a slow gain medium, such as QW lasers, due to the decoupling of light intensity and modal gain. For the QD lasers, their finite $\alpha_H$-factors give rise to a strong Kerr effect. On the other hand, the designed narrow ridge width of 2.6 $\mu$m also favors the Kerr nonlinearity, since the mode is more confined in the plane transverse to the propagation direction compared to a wide-ridge cavity \cite{gordon2008multimode}. Recently, we observed that the $\alpha_H$-factor of QD passively mode-locked laser can be increased by the SA reverse voltage \cite{dong2019frequency,dong20201}, indicating that the Kerr nonlinearity in the QD laser can be practically engineered. Figure \ref{Fig:3}c depicts the spectral $\alpha_H$-factors at the threshold of the QD FM comb laser under different SA reverse biases. With an increase in the SA voltage from 0 to -6 V, the $\alpha_H$-factor at gain peak increases from 1 to 3.1, which results from the decrease in differential gain and the increase in carrier density-induced change of the refractive index \cite{dong20201}. In this study, the $\alpha_H$-factors are measured by the amplified spontaneous emission (ASE) method, more details of the $\alpha_H$-factor measurements are available in the supplementary information. As a result of the increase in the $\alpha_H$-factor, a stronger Kerr nonlinearity takes place in the laser's active region. The data points near the gain peak, however, are missing because those modes suffer from a strong gain perturbation close to the threshold due to the SpaHB \cite{gordon2008multimode,bardella2017self}. This unfortunately prevents us from extracting the $\alpha_H$-factor of these modes accurately. In addition to the absolute value of the $\alpha_H$, the sign of the $\alpha_H$ and the Kerr nonlinearity is also crucial for the FM comb generation. The sign of the Kerr nonlinearity is simply determined by the slope of the spectral $\alpha_H$ through d$\alpha_H(\omega)$/d$\omega$ \cite{opavcak2021spectrally}, which is negative for the QD comb laser studied. 

It should be noted that examining the change in the Kerr effect alone does not explain the increase in the comb bandwidth, since the FM comb bandwidth is also dependent on the cavity dispersion. A theoretical study in QCLs revealed that the FM comb bandwidth is determined by a combination of the Kerr nonlinearity and the GVD \cite{opavcak2019theory}. The calibrated fiber interferometer method has been widely used to measure the GVD of high-Q cavities \cite{fujii2020dispersion}. Nevertheless, it is challenging to repeat this method in the low-Q FP cavities with high accuracy because the modal linewidth of each FP resonance is too broad to precisely locate the resonant frequencies. To address this issue, we propose a novel method for GVD measurements, the corresponding experimental setup is shown in Fig. \ref{Fig:3}a. This method allows for a simple and high-accuracy measurement after data processing (Methods). In this study, we measure the GVD of a single-section FP QD laser whose cavity is nearly identical to the one of the comb laser. The measured spectral first-order dispersion $\beta_1$ (blue dots) are plotted in Fig. \ref{Fig:3}b. The second-order dispersion $\beta_2$ (GVD) can thus be calculated by doing linear curve fit (blue curve) of the $\beta_1$ versus optical frequency. The measured average GVD over the selected spectral range is as low as 1200 fs$^2$/mm. In fiber optics, another widely used approach to analyze the GVD is the D-parameter. For the device studied, the corresponding D-parameter is found at -1333.5 ps/(km$\cdot$nm). Moreover, we investigate a simulation of the GVD to validate our method. Our calculation of GVD takes into account both the dispersion from the material and the cavity geometry, and the result (red curve) is shown in Fig. \ref{Fig:3}b. In the spectral range of the QD active region, the calculated GVD ranging from 1164 to 1257 fs$^2$/mm gives an average GVD of 1210 fs$^2$/mm, which is in good agreement with the experimental result. More details of the measurements and calculations of GVD are available in the supplementary information.

As a result of the negative Kerr nonlinearity and the normal GVD of the waveguide, an increase in the $\alpha_H$-factor contributes to compensation for the cold-cavity dispersion and thus a broadening of optical bandwidth. Such an effect has been predicted in a theoretical work for QCLs \cite{opavcak2019theory}. As one can see in Figs. \ref{Fig:2}, it is the Kerr nonlinearity that results in a one-fold improvement of the 3-dB comb bandwidth, as the SA reverse bias increases from -0.9 to -3.3 V. The contributions of SpaHB and FWM, however, are negligible since the gain current is fixed at 110 mA.

\medskip
\noindent \textbf{FM comb formation with four-wave mixing}

In single-section FM comb lasers, the gain current is found to largely improve the FM comb dynamics \cite{mansuripur2016single,dong2019frequency,chow2020multimode}. Such an effect is governed by the SpaHB and the FWM. The SpaHB refers to the fast carrier gratings with sub-wavelength spatial variation, which is generated by the optical field standing wave pattern in an FP cavity. As such, the gain of those cavity modes other than the primary mode is no longer clamped above the threshold but continues to increase with pumping. Therefore, SpaHB destabilizes the single-mode emission and triggers multimode instabilities. The SpaHB occurs simply by pumping the laser just a few percent above its threshold and is further strengthened with the increase in the gain current. On the other hand, the FWM originates from population pulsations created by the interaction of longitudinal modes of a semiconductor laser with the carrier reservoir. Temporal gain and index gratings are created, which result in the generation of new conjugate waves. After multimode emission occurs, the FWM contributes to the suppression of AM operation and phase locking of those lasing modes in time. It should be noted that the effectiveness of both the SpaHB and the FWM gratings is determined by the carrier recovery lifetime. Therefore, the FM combs are more likely to be triggered in a laser with a fast gain medium, such as QD lasers and QCLs, otherwise, the carrier gratings will be washed out by the carrier diffusion. In addition to the Kerr nonlinear effect discussed in the last section, the FWM is another driving force to compensate for the cold-cavity dispersion, which is beneficial for improving the comb bandwidth. 

\begin{figure*}[t!]
\centering
\includegraphics[width=\linewidth]{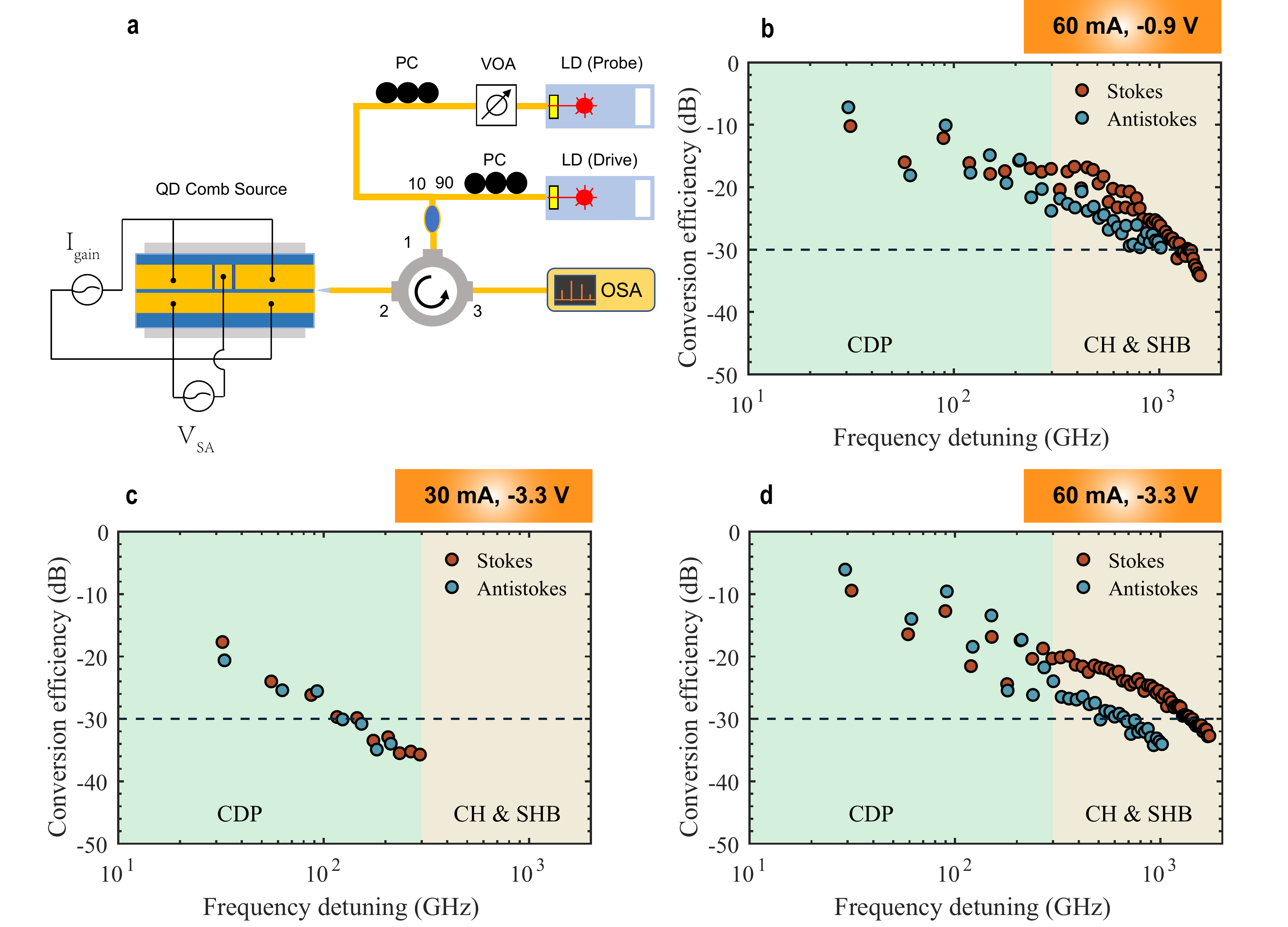}
\caption{\textbf{Four-wave mixing in QD FM comb laser.}
\textbf{a}. Experimental setup for the four-wave mixing (FWM) measurements. PC, polarization controller; VOA, variable optical attenuator; LD, laser diode; OSA, optical spectrum analyzer.
FWM efficiency as a function of frequency detuning, with the laser operation conditions ($I_{gain}$, $V_{SA}$) = 
\textbf{b}. (60 mA, -0.9 V);  
\textbf{c}. (30 mA, -3.3 V);
\textbf{d}. (60 mA, -3.3 V). CDP, carrier density pulsation; CH, carrier heating; SHB, spectral hole burning. 
}
\label{Fig:4}
\end{figure*}

The investigation of nondegenerate FWM in the QD comb laser studied is based on a pump-probe experimental setup, which is shown in Fig. \ref{Fig:4}a (Methods). While the QD laser is operating above the threshold, a drive laser (pump) is first injected into one longitudinal mode near the gain peak to lock the laser emission below the threshold. Another probe laser is then applied to one longitudinal mode other than the drive mode to generate the signal through the third-order nonlinear susceptibility $\chi^{(3)}$. In this study, the FWM conversion efficiency (CE) is defined as the ratio of the signal power $P_s$ to the probe power $P_p$, such as:
\begin{equation}
    \eta_{CE} = \frac{P_s}{P_p}
    \label{equ:CE}
\end{equation}
The CE thus gives an explicit description of the nonlinear conversion at the device level. It should be noted that the FWM efficiency in a semiconductor laser is also dependent on the frequency detuning between the probe and the drive waves $\Delta f$, which is a multiple of the cavity FSR. Depending on the frequency detuning, the FWM that results from population pulsations can be generated by carrier density pulsation (CDP), carrier heating (CH), and spectral hole burning (SHB). The effectiveness and bandwidth of those grating are determined by both the interband and the intersubband carrier capture and relaxation rates \cite{nielsen2010four}. Therefore, the ultrafast carrier recovery dynamics in QD allow for an efficient and broadband FWM process. With the increase in frequency detuning, the most efficient CDP grating plays a dominant role in the low detuning range (0 - $\sim$100 GHz); the CH and the SHB gratings will contribute to a broadband FWM operation over 1 THz. More details about the FWM process in QD laser are available in the supplementary information. The frequency detuning is further classified into stokes (frequency up-conversion) and antistokes (frequency down-conversion), depending on if the generated signal's frequency $f_{signal}$ is higher or lower than the probe's frequency $f_{probe}$, such as:
\begin{equation}
  \Delta f = f_{probe} - f_{drive} 
  \begin{cases}
      < 0 & \text{(Stokes)}\\
      > 0 & \text{(Antistokes)}
    \end{cases}   
\label{equ:FWM_detuning}
\end{equation}
where $f_{drive}$ denotes the drive wave frequency. The coincidence between the stokes and the antistokes operation is determined by the symmetry of the gain profile, which depends on the $\alpha_H$-factor.

\begin{table*}[t!]
\begin{center}
\caption{Comparison of Recent QD Mode-Locked Laser on Various Material Platforms and Structures.}
\label{tab:compare}
\begin{tabular}{|p{0.07\linewidth}|p{0.15\linewidth}|p{0.19\linewidth}|p{0.06\linewidth}|p{0.06\linewidth}|p{0.08\linewidth}|p{0.15\linewidth}|p{0.08\linewidth}|p{0.08\linewidth}|}
\hline
Telecom band & Material platform$^{\dagger}$ & Method & \multicolumn{2}{c|}{Pulse width (ps)} & Repetition rate (GHz) & Optical bandwidth (nm) & Operation temperature ($^{\circ}$C) & Ref. \\
\hline
&&& AM & FM &&&&\\
\hline 
C-band & InAs/InP & Single-section & & 0.295 & 10-100 & 17.9 (-3 dB BW) $^{\ddagger}$ & 18 & \cite{lu2011ultra} \\
C-band & InAs/InP & Single-section & & 0.6 & 34.2 & 11.96 (-6 dB BW) & 18 & \cite{lu2021inas}\\
C-band & InAs/InP & Single-section & & & 32.5 & 11.8 (-3 dB BW) & 25 & \cite{verolet2020mode}\\
O-band & InAs/InGaAs on Si & Single-section & & 0.49 & 31 & & 20 & \cite{liu2018490} \\
O-band & InAs/GaAs on Si	& Two-section & 5 & & 20 & 6.1 (-3 dB BW) & 18 & \cite{liu2019high} \\
O-band & InAs/GaAs & Two-section & 4.9 & & 25.5 & 4.7 (-6 dB BW) & 20 - 120 & \cite{pan2020quantum} \\
O-band & InAs/GaAs & CPM & 0.81 & & 100 & 11.5 (-3 dB BW) & 25 - 100 & \cite{huang2022ultra} \\
O-band & InAs/GaAs on SOI & CPM (external cavity) & & & 102 & 6.5 (-3 dB BW) & 25 &  \cite{kurczveil2018chip}\\
O-band & InAs/GaAs on SOI & CPM (external cavity) & & & 15.5 & 12 (-3 dB BW) & 23 & \cite{liang2022energy} \\
O-band & InAs/GaAs & CPM & 1.7 & 0.495 & 60 & 12.1 (-3 dB BW) & 20 - 45 & This work \\
\hline
\end{tabular}
$^{\dagger}$ A performance comparison that includes the QW lasers are available in the Supplementary Information.\\
$^{\ddagger}$ A span of 17.9 nm in the C-band approximates 2.2 THz, which is on par with the bandwidth that our QD laser delivers.
\end{center}
\end{table*}

Here, we begin with an analysis of the influence of gain current on the FWM efficiency. With the SA reverse bias fixed to -3.3 V, the FWM conversion efficiencies as a function of frequency detuning when the gain current is 30 mA and 60 mA are shown in Figs. \ref{Fig:4}c and d, respectively. The stokes and antistokes operations are presented by the red and the green markers, respectively. Our results demonstrate that the FWM efficiency in the full range of frequency detuning is improved by approximately 10 dB, as the gain current increases from 30 to 60 mA. If a low current such as 30 mA is applied to the gain section, the CEs are too low to be measured when the frequency detuning is larger than 300 GHz. This weak FWM results from the slow gain dynamics since the wetting layer and the excited states are not filled up. The maximum CE as large as -5 dB is found at 60 mA, where the frequency detuning is the first cavity FSR at 30 GHz. With the increase in frequency detuing from 30 to 300 GHz, the FWM efficiency decreases from -5 to $\sim$ -20 dB until it is quasi-constant between 300 and 500 GHz. Such a plateau of CE is attributed to the CH and SHB gratings. On one hand, the contribution of CDP to the FWM efficiency still plays a dominant role when the frequency detuning is as large as 300 GHz, which is ten times the cavity resonant frequency. On the other hand, the effectiveness of CH and SHB gratings still exists as the frequency detuning increases to 1.7 THz, which contributes to the -30-dB FWM bandwidth as large as 1.4 THz in the stokes case. As such, the broadband FWM grating confirms that the carrier recovery rate in QD is fast enough for FM comb generation. As a result of the improved FWM, a higher gain current contributes to the suppression of AM operation and the formation of a broadband FM comb. As one can see in Fig. \ref{Fig:AM_FM}, the QD laser experiences a transition from AM dominant operation to FM dominant operation as the gain current increases. It should be noted that an FWM efficiency as large as -5 dB is comparable to a single-section QD laser, which is more than 20 dB higher than a QW laser \cite{duan2022four}. 

We end this section with an analysis of the influence of SA reverse bias on FWM efficiency. When the QD laser operates with a gain current at 60 mA, the QD laser does not suffer from any degradation of FWM efficiency in the stokes condition, as the SA reverse bias increases from -0.9 to -3.3 V. As such, the gain dynamics in the QD laser is still fast enough for FM comb generation. Nevertheless, the FWM bandwidth in the antistokes condition is reduced by the SA reverse bias, which results in a discrepancy between the stokes and antistokes operations at -3.3 V. It should be noted that the symmetry of stokes and antistokes FWM is dependent on the gain profile. When a low voltage is applied to the SA, the low $\alpha_H$-factor contributes to a symmetric distribution of modal gain with respect to the drive photon frequency. Therefore, a pair of generated signal photons with the same frequency detuning can be amplified by a similar amount. Once the $\alpha_H$-factor is increased by the SA reverse bias, the gains of those signal photon pairs are different, which results in their different conversion efficiencies. In both SA bias conditions, the FWM efficiencies that are determined by the CDP exhibit a clear undulation against the variation of frequency detuning, which is rarely observed in single-section semiconductor lasers and semiconductor optical amplifiers (SOAs). Such an effect is possibly caused by the different overlap factors for the gain and the SA. When the QD laser is injection-locked by the drive laser, the modes that are separated by an odd number of FSR from the primary mode have larger gains than the even number-spaced modes. As a result, their FWM efficiencies are higher. Detailed discussions of the overlap factors in the colliding-pulse mode-locked laser are available in the Supplementary Information.

\medskip
\begin{large}
\noindent \textbf{Discussion} 
\end{large}

In this work, we report a 60 GHz QD mode-locked laser that allows for the generation of both AM and FM combs. In-depth analyses of the underlying physics are investigated to improve the laser performance. We demonstrate that semiconductor QDs are promising for AM/FM comb generation, owing to their three-dimensional charge carrier confinement and discrete energy levels that allow for both slow and fast gain dynamics. In addition to the engineering of the GVD of the waveguide, we demonstrate that the Kerr nonlinearity can be practically engineered to improve the FM comb bandwidth, which gives novel insights into the development of FM combs. Compared to the conventional AM comb, the broadband nature of FM combs is beneficial to high-capacity optical communication system. The performance comparison of QD mode-locked lasers on various designs is presented in Tab. \ref{tab:compare}. Our QD platform allows for fabricating devices whose 3-dB bandwidth exceeds 2.2 THz (12.1 nm), which is on par with the best QD mode-locked lasers reported so far. Both the AM and the FM pulse widths generated in our devices meet the state-of-art of the QD mode-locked laser. Different from previous work, our approach allows for AM/FM comb generation from a single device. This broadband dual-mode mode-locked laser thus opens up new opportunities for spectroscopy, remote sensing, and optical frequency synthesis. Despite the demonstration of FM combs in conventional QW lasers, their weak SpaHB, FWM, and Kerr effect prevent them to be efficient FM comb generators. In this context, the QD laser is a compelling solution for FM comb sources in the near-infrared spectral range. Other FM comb platforms such as QCLs and ICLs would possibly benefit from the optimization path proposed in this work to improve laser performance.

Compared to the other solutions for integrated OFC technologies, the advantages of QD FM comb lasers come from their robustness, high energy efficiency, simple design and fabrication, and low-cost thanks to their compatibility with mature complementary metal–oxide–semiconductor very-large-scale integration (CMOS-VLSI) fabrication technology \cite{liu2019high}. Compared to the Kerr comb and electro-optics (EO) comb \cite{Jin2021hertz,xiang2021laser}, the QD laser comb solution offers a broadband and flat-topped comb without the need for careful engineering of the waveguide dispersion, which facilitates the device design and fabrication. It should be noted that the comb bandwidth offered by the QD laser strikes a good balance between the wall-plug efficiency and the performance of a DWDM system in which many microring modulators are deployed. Therefore, our approach is being pursued by both academia and industry. Despite the higher frequency noise than its counterpart, the QD laser comb is still a compelling solution for high-volume data transmission. The transmission capacity of a single comb line in a QD laser comb is beyond 40 Gbaud PAM4 \cite{huang2022ultra}, which allows for a total transmission capacity exceeding 12 Tbps \cite{kemal202032qam,liu2023mode}. Linewidth locking technologies are being investigated to further reduce the QD laser linewidth to kilohertz level and sub-kilohertz level \cite{verolet2020mode}.

\medskip
\begin{large}
\noindent \textbf{Methods} 
\end{large}

\medskip
\noindent \textbf{QD FM comb laser fabrication}

The samples were grown with a Varian GenII Molecular Beam Epitaxy chamber on nominal 001 n++ GaAs substrates, and the layer structure is shown in Fig. \ref{Fig:SEM}(b). The waveguide consists of a 330 nm active region between 1.4 $\mu$m 40\% AlGaAs n- and p-type cladding layers and 300 nm highly doped contact layers. The aluminium content is linearly graded between the cladding and the contact layers and the 20\% AlGaAs confinement layers.  The active region contains 6 QD layers embedded in 7 nm 15\% InGaAs quantum wells, commonly referred to as dots in a well (DWELL). The wafer was processed into shallow etched lasers by dry etching ridges, which can be seen in Fig. \ref{Fig:SEM}(a). Additionally, the laser underwent a second etch to electrically isolate the gain and SA regions by etching through the p-contact and 300 nm of the p-cladding.

\medskip
\noindent \textbf{FM comb characterizations}

The QD laser is placed on a temperature-controlled stage, with a test temperature set to 20 $^{\circ}$C throughout the whole measurements. The laser emission is coupled by a polarization-maintained (PM) fiber and then passes through a fiber optical isolator before it is further analyzed. The optical spectra are measured by a high-resolution optical spectrum analyzer (OSA) (Yokogawa AQ6370C). After the laser emission is captured by a high-speed photodiode (PD) (Finisar HPDV2120R), the 60 GHz RF beatnote is firstly down-converted to 20 GHz by using a low-noise RF local oscillator (LO) (Agilent E8257D) through an RF mixer (Marki Microwave MM1-2567LS). The RF signal is then measured by an electrical spectrum analyzer (ESA) (Rohde \& Schwarz FSU50). The laser is pumped by a low-noise laser current source (ILX Lightwave LDX-3620) to ensure a stable and low-noise operation. The intensity autocorrelation is firstly generated by a high-sensitivity autocorrelator (ATC) (Femtochrome FR-103MN) before it is measured by a high-resolution real-time oscilloscope (Agilent MSO7104B). 

\medskip
\noindent \textbf{Intermode beating phase measurement}

After the FM comb operation is well developed, an external low-noise tunable laser (TNL) (Keysight 81606A) will beat with the m$^{th}$ and the (m+1)$^{th}$ comb lines through a 50/50 fiber beam splitter. The frequency detuning between the TNL and the m$^{th}$ mode $\delta$, and the frequency detuning between the TNL and the (m+1)$^{th}$ mode $f_{rep}-\delta$ are then generated, where $f_{rep}$ is the 60 GHz laser repetition frequency. Both these two RF beat signals as well as the laser beatnote are then captured by a high-speed photodiode (PD) (Finisar HPDV2120R). All these signals are amplified by a 30-dB RF amplifier (AMP) (RF-Lambda RLNA00G18GA) before they are analyzed by a high-speed real-time oscilloscope (Keysight UXR0594A). More details on the data treatment are available in the supplementary information.

\medskip
\noindent \textbf{Group velocity dispersion measurement}

The FP laser investigated in this measurement has an identical volume as the QD mode-locked laser. The cavity length of 1.35 mm yields a fundamental cavity resonance of 30 GHz. Laser emissions from both facets of the QD FP laser are coupled by AR-coated lens-end single-mode fibers. After the laser is turned off, the emission of a semiconductor optical amplifier (SOA) (Thorlabs S9FC1132P) is sent into the FP cavity from the front facet to generate an interferometric fringe pattern. The output light from the rear facet is then measured by a high-resolution optical spectrum analyzer (OSA) (Yokogawa AQ6370C). Detailed data processing is explained in the supplementary information.

\medskip
\noindent \textbf{Four-wave mixing experiment}
 
While the laser is operating above the threshold, a low-noise tunable laser (drive laser) (EXFO T100S-HP) is applied to one longitudinal mode at the gain peak to lock the laser emission below the threshold. Another low-noise tunable laser (probe laser) (Keysight 81606A) is then applied to another longitudinal mode other than the drive frequency to generate the signal. While the output powers of the drive and the probe lasers are fixed to 10 dBm, a variable optical attenuator (VOA) (Keysight 81570A) is used to finely change the probe power. The FWM spectra are measured by a high-resolution optical spectrum analyzer (OSA) (Yokogawa AQ6370C).

\medskip
\begin{footnotesize}


\noindent \textbf{Acknowledgments}: 
This work is supported by the Defense Advanced Research Projects Agency (DARPA) PIPES programs. B. D. acknowledge Nikola Opačak and Benedikt Schwarz from TU Wien, Austria; Jianan Duan from Harbin Institute of Technology, Shenzhen, China, for fruitful discussions.

\noindent \textbf{Author contribution}: 
B.D. characterized and gathered the experimental data from the device, including the FM comb dynamics characterizations, the linewidth enhancement factor, the waveguide group velocity dispersion, and the four-wave mixing, with contributions from M.D., O.T., and A.N. 
M.D. designed and fabricated the QD devices.
H.W. provided theoretical calculations and analysis on the waveguide group velocity dispersion.
B.D. wrote the manuscript with inputs from M.D., and H.W.. 
All authors commented on and edited the manuscript.
J.E.B. supervised the project.


\noindent \textbf{Data Availability Statement}: 
All data generated or analyzed during this study are available within the paper and its Supplementary Information. Further source data will be made available upon reasonable request.

\end{footnotesize}

\bibliographystyle{apsrev4-2}

%

\pagebreak

\end{document}